\documentstyle[sprocl]{article}

\input{epsf}
\def\Journal#1#2#3#4{{#1} {\bf #2}, #3 (#4)}

\def\NPB{{\em Nucl. Phys.} B}

\def\PRD{{\em Phys. Rev.} D}

\def\be{\begin{equation}}
\def\ee{\end{equation}}
\def\bea{\begin{eqnarray}}
\def\eea{\end{eqnarray}}

\newcommand{\bc}{\begin{center}}
\newcommand{\ec}{\end{center}}
\newcommand{\beqn}{\begin{equation}}
\newcommand{\eeqn}{\end{equation}}
\newcommand{\barr}{\begin{eqnarray}}
\newcommand{\earr}{\end{eqnarray}}

\begin{document}

\title{Instanton, Monopole and Confinement}

\author{S.~Sasaki \footnote[2]{Talk presented by S.~Sasaki
(E-mail address: ssasaki@rcnp.osaka-u.ac.jp)}, 
M.~Fukushima, A.~Tanaka, H.~Suganuma, H.~Toki,\\
O.~Miyamura$^{\;*}$ and D. ~Diakonov$^{\;**}$}

\address{Research Center for Nuclear Physics, Osaka University, 
Osaka 567, Japan \\
Department of Physics, Hiroshima University,
Higashi-Hiroshima 739, Japan$^{\;*}$\\
Nuclear Physics Institute, St. Petersburg 188350, 
Russia$^{\;**}$}


\maketitle\abstracts{
We study the correlation between instantons and QCD-monopoles
both in the lattice gauge theory and in the multi-instanton system
using the maximally abelian gauge.
First, we find the existence of an almost linear
correlation between the total length of monopole trajectories and
the total number of pseudoparticles (instantons and anti-instantons)
in the $16^{3}\times4$ 
SU(2) lattice.
Second, we study the features of QCD-monopole in
the SU(2) multi-instanton vacuum on the $16^{4}$
lattice as a random ensemble of pseudoparticles.
A signal of monopole condensation is found
as the clustering of monopole trajectories,
when the topological pseudoparticles is sufficiently dense.
}

\section{Topological Objects in QCD}
\indent

The appearance of QCD-monopole was pointed out 
by 't Hooft using abelian gauge fixing \cite{thooft}. 
Lattice QCD calculations \cite{{kron},{pol}} show that 
this topological object
plays an essential role on color confinement through its
condensation, which is characterized by
the appearance of the large monopole clusters and can be interpreted as a
Kosterlitz-Thouless-type phase transition \cite{ezawa}.
In QCD, there is also another non-trivial topological object, 
a instanton.
Instantons and QCD-monopoles are thought to be
hardly related to each other
since these topological objects appear from
different non-trivial homotopy groups.
Recently, however, the existence of a relation between the two objects 
has been shown by analytic studies and
lattice QCD simulations \cite{{pol},{suga},{mar}}.

In our previous analytical works, we conjectured that
the existence of instantons 
promotes the clustering of monopole trajectories
as a signal of monopole condensation \cite{suga}.
First, we examine an evidence for our conjecture
by measuring the monopole-loop
length and the number of instantons simultaneously in the lattice QCD 
simulation. For this discussion, we take the SU(2) gauge theory.

\section{Lattice Study for Monopole Trajectory and Instantons}
\indent

We study the correlation between the total monopole-loop
length $L_{\rm total}$ and the integral of the absolute value of the 
topological density $I_{Q}$, which corresponds to the total number
of instantons and anti-instantons, by use of the Monte Carlo 
simulation in the lattice gauge theory of the SU(2) Wilson action.
We can measure $L_{\rm total}$ and $I_{Q}$ using the following procedure.

In the maximally abelian (MA) gauge,
the abelian gauge fixing is done by maximizing
 $R \equiv \sum_{s, \mu} {\rm Tr}\{ U_{\mu}(s) \tau^{3}
U_{\mu}^{-1}(s) \tau^{3}\}$.
The SU(2) link variable $U_{\mu}(s)$ is then factorized into the 
abelian link variable; $u_{\mu}(s)=\exp \{ i\tau_{3}\theta(s) \}$
and off-diagonal part; $M_{\mu}(s)\equiv\exp \{ i\tau_{1}
C^{1}_{\mu}(s)+i\tau_{2}C^{2}_{\mu}(s) \}$ as 
$U_{\mu}(s)=M_{\mu}(s)u_{\mu}(s)$.
The Dirac string is extracted from
the abelian field strength $\theta_{\mu \nu}\equiv
\partial_{\mu}\theta_{\nu}-\partial_{\nu}\theta_{\mu}$
by decomposition as
$\theta_{\mu \nu}={\bar \theta}_{\mu \nu}+2\pi
M_{\mu \nu}$ with $-\pi \leq{\bar \theta}_{\mu \nu}< \pi$ and
$M_{\mu \nu}\in {\bf Z}$. Here, ${\bar \theta}_{\mu \nu}$ and
$M_{\mu \nu}$ correspond to the regular part and the Dirac
string part, respectively \cite{{kron},{suga}}.
The monopole current is derived from the Dirac string part as
%
%
\beqn
k_{\mu}(s)={1 \over 2}
\varepsilon_{\mu \nu \rho \sigma}\partial_{\nu}M_{\rho \sigma}
(s+\hat \mu)\;\;\;,
\eeqn
which forms a closed loop, since the monopole current is 
topologically conserved.
The total monopole-loop length is measured as
$L_{\rm total}=\sum_{\mu, s}\mid k_{\mu}(s)\mid$. 
In order to examine the total number
of topological pseudoparticles, 
the integral of the absolute value of the 
topological density is defined as 
%
%
\beqn
I_{Q}={1 \over 32\pi^{2}}\sum_{s}
\varepsilon_{\mu \nu \rho \sigma}
\mid{\rm Tr} \{ U_{\mu \nu}(s)U_{\rho \sigma}(s) \}\mid
\eeqn
where $U_{\mu \nu}(s)$ is the plaquette variable.
This value is measured with the Cabibbo-Marinari cooling method
in the same way as the topological charge.

We make lattice calculations for various $\beta=2.2\sim 2.35$ on the
$16^{3}\times 4$ lattice,
where the deconfinement transition occurs at $\beta_{c}\simeq 2.3$.
As shown in Fig.1, we find the almost linear correlation between
$I_{Q}$ after 3 cooling sweeps
and $L_{\rm total}$.
Hence, the monopole-loop length would be largely 
enhanced in the dense instanton system.

\section{Clustering of Monopoles in the Multi-Instanton System}
\indent

We study the multi-instanton system by measuring the monopole clustering
in the abelian gauge
in order to understand the role of instantons on confinement 
\cite{fukushima}.
The field configuration 
for single instanton with the size $\rho$ and the center $z_{\mu}$
in the singular gauge is
%
%
\beqn
A^{I}_{\mu}(x;z_{k},\rho_{k},O_{k})=
{{i\rho^{2}\tau_{a}O^{ai}{\bar \eta}^{i}_{\mu \nu}(x-z)_{\nu}}
\over {(x-z)^{2}[(x-z)^{2}+\rho^{2}]}}\;\;,
\eeqn
where $O^{ai}$ is the color orientation matrix and 
${\bar \eta}^{i}_{\mu \nu}$ the 't Hooft symbol.
For {\it anti}-instantons $A^{I}_{\mu}(x;z_{k},\rho_{k},O_{k})$,
one has to replace the ${\bar \eta}^{i}_{\mu \nu}$ symbol by 
$\eta^{i}_{\mu \nu}$.
The multi-instanton configurations are assumed
as the sum of instanton ($I$) and
anti-instanton ($\bar I$) solutions,
%
%
\beqn
A_{\mu}(x)=\sum_{k}A^{I}_{\mu}(x;z_{k},\rho_{k},O_{k})
+\sum_{k}A^{\bar I}_{\mu}(x;z_{k},\rho_{k},O_{k})\;\;\;.
\label{suminst}
\eeqn 
We generate ensembles of $N$ pseudoparticles with random orientations 
and random positions. The size of pseudoparticles are taken 
according to the size distribution $f(\rho)$.
Here, we adopt the following instanton size distribution 
\cite{fukushima} as
%
%
\beqn
f(\rho)={1 \over {{\left({\rho \over \rho_{_{\rm IR}}}\right)^{\nu}}
		+{\left({\rho_{_{\rm UV}} \over \rho}\right)^{b-5}} } }
\label{dis}
\eeqn
with $b = {11 \over 3}N_{c}$.
Here, $\rho_{_{\rm IR}}$ and $\rho_{_{\rm UV}}$ are certain parameters 
such that
the distribution (\ref{dis}) is normalized to unity, while the 
maximum of the distribution is fixed to a give value $\rho_{0}$,
which is the most probable size of the pseudoparticles in the ensemble.
It is noted that $f(\rho)$ is reduced to the one 
obtained in the dilute instanton case; $f(\rho)\sim
\rho^{b - 5}$ in the limit of $\rho \rightarrow 0$.
For large size instantons, $f(\rho)$ falls off with negative power of 
$\nu$,
since large size instantons are suppressed by the instanton interaction.
Here, we adopt $\nu=3$, which is proposed 
by Diakonov and Petrov \cite{fukushima}.
We then introduce a lattice and express the gauge field in terms 
of the unitary matrices $U_{\mu}=\exp(iaA_{\mu})$ living on the links.
Then, we can do exactly the same procedure as done in lattice 
simulations in order to extract the monopole trajectories in the MA gauge.

We take a $16^{4}$ lattice with the lattice spacing of $a=0.15 {\rm 
fm}$ and the most probable instanton size, $\rho_{0}=0.4 {\rm fm}$. 
The volume is thus fixed and equal to $V=(2.4{\rm fm})^{4}$.
In this calculation, we take equal numbers of instantons and anti-instantons; 
$N_{I}=N_{\bar I}=N/2$.
In Fig.2, we show the histograms of monopole
loop length for two typical cases with the total pseudoparticle 
number $N=20$ and 60, which  correspond to the density 
$(N/V)^{1 \over 4}=174$ and 
$228 {\rm MeV}$, respectively.
At low instanton density, only relatively short monopole loops are found.
At high density, there appears one very long monopole loop in {\it 
each} gauge configuration. 

\section{Summary}
\indent

From above results, 
instantons would be the source of large size monopole clustering,
which indicates occurring of monopole condensation
in the similar argument as the Kosterlitz-Thouless-type phase 
transition \cite{ezawa}.
Thus, instantons seem to play a relevant role on color confinement.

%
%
\begin{figure}
\centerline{\epsfxsize=2.8in 
\epsfbox{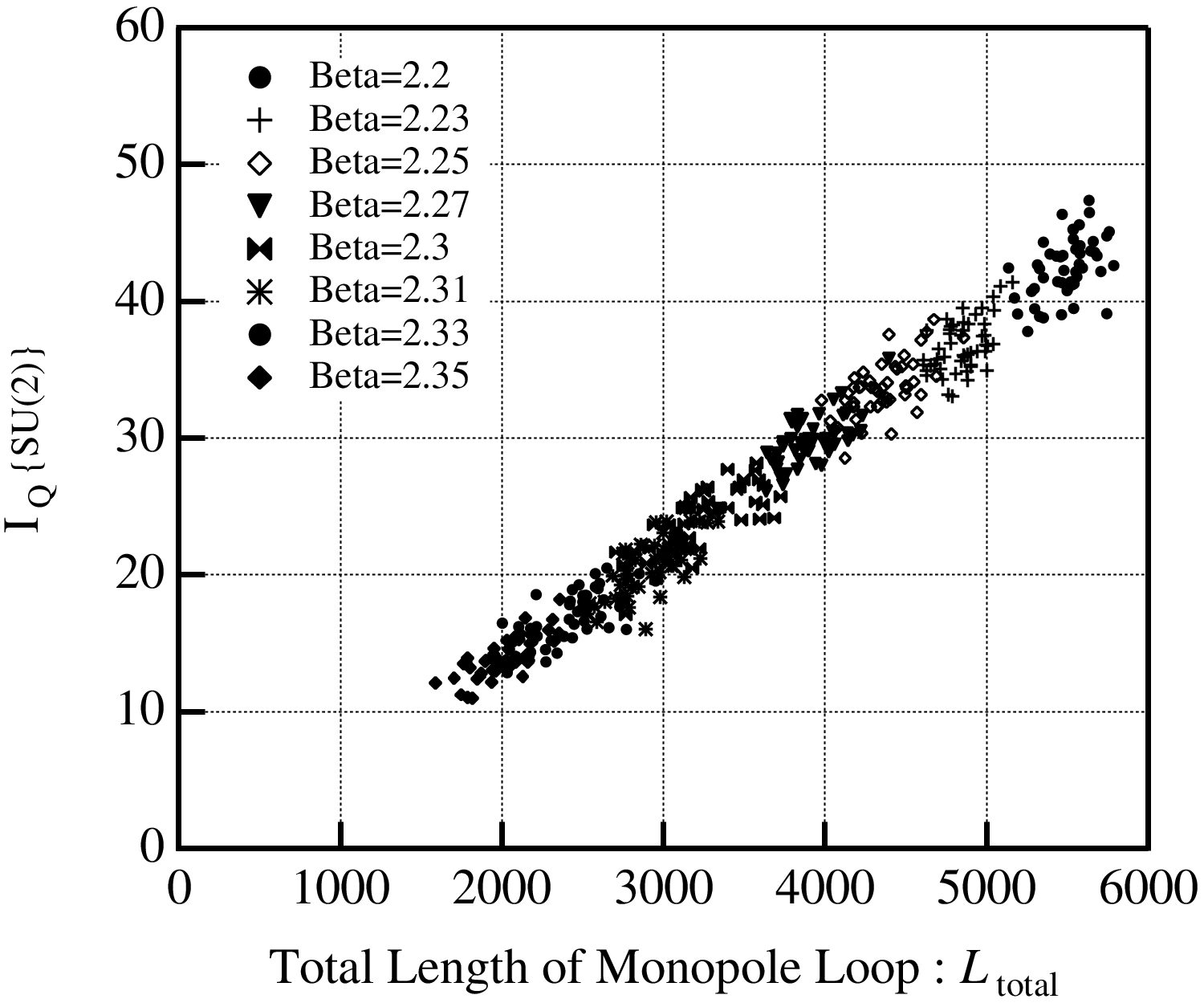}}
{\small Fig.1~Correlation between 
$L_{\rm total}$ and $I_{Q}$
in MA gauge on $16^{3}\times
4$ lattice with various $\beta$.
We use 3 cooling sweeps for the calculation of $I_{Q}$.}

\vspace*{0.3cm}

\newlength{\minitwocolumn}
\setlength{\minitwocolumn}{2.2in}
\begin{minipage}{\minitwocolumn}
\centerline{\epsfxsize=2.0in 
\epsfbox{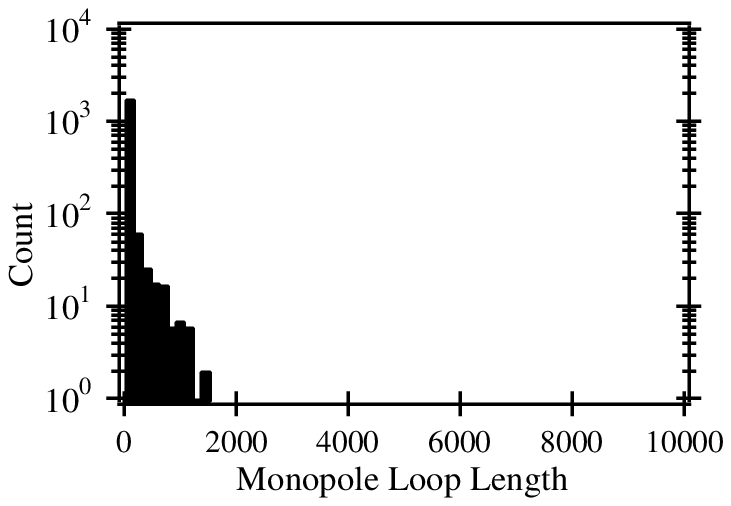}}
\end{minipage}
\hspace{\columnsep}
\setlength{\minitwocolumn}{2.2in}
\begin{minipage}{\minitwocolumn}
\centerline{\epsfxsize=2.0in 
\epsfbox{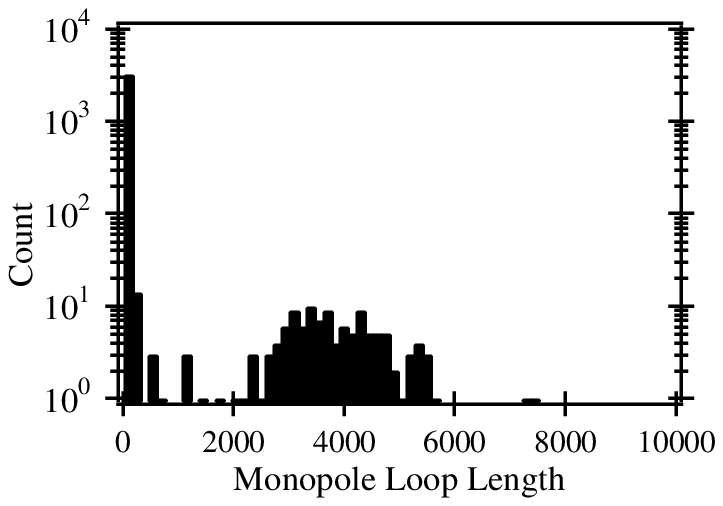}}
\noindent
\end{minipage}
{\small
Fig.2~Histogram of monopole loop length in the multi-instanton
system; (a) dilute case ($N=20$) and (b) dense case ($N=60$).}
\end{figure}

\end{document}